\documentstyle[preprint,aps,epsfig]{revtex}
\def\bge{\begin{equation}}
\def\ene{\end{equation}}
\def\bg{\begin{eqnarray}}
\def\en{\end{eqnarray}}
\def\nn{\nonumber}

\def\bi{\bibitem}
\begin{document}
\preprint{
\vbox{
\hbox{Invited talks presented at the Workshop on}
\hbox{Quarks Hadrons and Nuclei,}
\hbox{Adelaide, November 1995.}
\hbox{ADP--96--7/T212}
}}
\title{Quarks in Finite Nuclei}

\author{{\it P. A. M. Guichon$^{\rm a,d}$, 
             K. Saito$^{\rm b,d}$}, 
    {\it and A. W. Thomas$^{\rm c,d}.$}\\
 $^{\rm a}$ DAPHIA/SPhN, CE Saclay, \\
          91191 Gif-sur-Yvette,\\ 
          CEDEX, France. \\
 $^{\rm b}$ Physics Division, Tohoku College of Pharmacy, \\
          Sendai 981, Japan. \\
 $^{\rm c}$ Department of Physics and Mathematical Physics, \\
          University of Adelaide, \\
          S.A. 5005 Australia. \\
 $^{\rm d}$ Institute for Theoretical Physics,
		University of Adelaide, \\
		  S.A. 5005 Australia.} 

%
\maketitle

\newpage
\noindent{\it Abstract:}

We describe the development of a theoretical description of the
structure of finite nuclei based on a relativistic quark model of the
structure of the bound nucleons which interact through the (self-consistent) 
exchange of scalar and vector mesons.
 
\section{Introduction}

By now it is well established that one needs many-body forces to
understand the structure of atomic nuclei. There are many ways of
dealing with this problem. In the space available we cannot review the
problem in general, rather we shall focus on recent progress based on
one specific model -- the quark-meson coupling model originally proposed by
Guichon\cite{guichon}.

The quark-meson coupling
model may be viewed as an extension of QHD\cite{walec} in which the
nucleons still interact through the exchange of $\sigma$ and $\omega$ 
mesons. However, the mesons couple not to point-like nucleons but to
confined quarks. In studies of infinite nuclear matter it was found that
the extra degree of freedom provided by the internal structure of the
nucleon means that one gets quite an acceptable value for the
incompressibility once $g_{\sigma}$ and $g_{\omega}$ are chosen to
reproduce the correct saturation energy and density. This is a
significant improvement on QHD\cite{walec,serot}
at the same level of sophistication.

In the light of current experimental work in relativistic heavy ion
collisions, which produce nuclear matter at densities several times
normal, there has been some initial work on the variation of baryon and
meson properties with density using the quark-meson coupling
model\cite{hadrons}.
There have also been some interesting applications to the properties of
finite nuclei using the local-density approximation, notably the
Okamoto-Nolen-Schiffer anomaly\cite{ons} and super-allowed Fermi 
$\beta$-decay\cite{wilk}.
However, the inherent problems of the local-density approximation mean
that these applications can at best be semi-quantitative and it is
clearly very important that the extension to finite nuclei be developed.

Our aim here is review a recently developed formulation of the
quark-meson coupling 
model for finite nuclei\cite{finite}, based on the Born-Oppenheimer
approximation. We shall pay particular attention to the spin-orbit force
in the model and its relation to the corresponding force in conventional
models involving meson exchange between point-like nucleons. Some
initial results for finite nuclei will also be presented.

\section{The Born-Oppenheimer Approximation for Finite Nuclei}

The solution of the general problem of a composite, quantum particle
moving in background scalar and vector fields that vary with position is
extremely difficult. One has a chance to solve the particular problem of
interest to us, namely light quarks confined in a ``nucleon" which is
itself bound in a finite nucleus, only because the nucleon motion is
relatively slow and the quarks highly relativistic. Thus the 
Born-Oppenheimer approximation, in which the ``nucleon" internal
structure has time to adjust to the local fields, is naturally suited to
the problem. It is relatively easy to establish that the method should
be reliable at the level of a few percent.

Even within the Born-Oppenheimer approximation the nuclear surface gives
rise to external fields that may vary appreciably across the finite size
of the nucleon. Our approach has been to start with a 
classical ``nucleon"
and to allow its internal structure (quark wavefunctions and bag radius)
to adjust to minimise the energy of three quarks in the ground-state of
a system consisting of the bag plus constant scalar and vector fields,
with the values at the centre of the ``nucleon". (From now on we shall
not put quotation marks on ``nucleon", but it should be remembered that
our bound nucleon is a quasi-particle whose structure necessarily
differs from that of a free nucleon.) Of course, the major problem with
the MIT bag (as with many other relativistic models of nucleon
structure) is that it is difficult to boost. We therefore solve the bag
equations in the instantaneous rest frame (IRF) of the nucleon -- using a
standard Lorentz transformation to find the energy and momentum of the
classical nucleon bag in the nuclear rest frame.

Having solved the problem using the fields at the centre of the nucleon
one can then use perturbation theory to correct for the variation of the
scalar and vector fields across the bag. In first order perturbation
theory only the spatial components of the vector potential 
give a non-vanishing 
contribution. (Note that although in the nuclear rest frame only the
time component of the vector field is non-zero, in the nucleon IRF there
are also non-vanishing spatial components.) 
This extra term is a small spin orbit correction to the
energy 
\begin{equation}
\delta M_N^{\star}(\vec{R}) = \eta_s(\vec{R})
\frac{\mu_s}{M^{\star 2}_N(\vec{R}) R} \left( \frac{d}{dR}
3 g_\omega^q\omega(\vec{R}) \right) \vec{S} \cdot\vec{L},
\label{1}
\end{equation}
where $\mu_s$ is the isoscalar magnetic moment of the nucleon bag, 
$3 g_\omega^q\omega$ is the vector potential felt by the nucleon with
effective mass $M^{\star}_N$ and $\eta_s$ is a correction factor of order
unity. In retrospect it is not surprising that the scalar magnetic
moment appears, as this correction is associated with the effective
magnetic field of the vector potential.

The interaction in Eq.(\ref{1}) induces a rotation of the spin as a
function of time. However, even if
$\mu_s$ were equal to zero, the spin would rotate because of
Thomas precession. Suppose that at time $t$, the spin vector is
$\vec{S}(t)$ in the  IRF($t$).
Then we expect that, at time $t+dt$ the spin
has the same direction if it is viewed from the frame
obtained by boosting the IRF($t$) by $d\vec{v}$ so as to get the right
velocity $\vec{v}(t+dt)$. That is, the spin looks at rest in the frame
obtained
by first boosting the NRF to $\vec{v}(t)$ and then boosting by
$d\vec{v}$.
This product of Lorentz transformation amounts to a boost to
$\vec{v}(t+dt)$
times a rotation. So, viewed from the IRF($t+dt$), the spin appears to
rotate. In order that our Hamiltonian be correct it should contain a
piece $H_{prec}$ which produces this rotation through the Hamilton
equations of motion. A detailed derivation can be found in
Refs.\cite{jackson,goldstein} and the result is
\bge
H_{prec}=-\frac{1}{2}\vec{v}\times\frac{d\vec{v}}{dt} \cdot \vec{S}.
\label{2}
\ene

One may find the acceleration corresponding to the interaction (\ref{2})
from the Hamilton equations of motion. This gives
\bge
\frac{d\vec{v}}{dt}=-\frac{1}{M_N^{\star}(\vec{R})}
\vec{\nabla}[M_N^{\star}(\vec{R})+3g_\omega^q\omega(\vec{R})].
\label{3}
\ene
If we put this result into Eq.(\ref{2}) and add the result  to
Eq.(\ref{1}),  we get the total spin orbit interaction (to first order
in the velocity)
\bge
H_{prec.} + H_1 = V_{s.o.}({\vec R}) \vec{S} \cdot\vec{L},
\label{4}
\ene
where
\bge
V_{s.o.}({\vec R}) = -\frac{1}{2M^{\star 2}_N(\vec{R}) R}
\left[ \left( \frac{d}{dR} M^{\star}_N(\vec{R}) \right)
+(1-2\mu_s\eta_s(\vec{R})) \left( \frac{d}{dR}
3g_\omega^q\omega(\vec{R})
\right) \right].
\label{5}
\ene

\subsection{Centre of Mass Motion}

We have already mentioned the difficulty of boosting the bag, a problem
which is closely related to the removal of spurious centre of mass
motion. In Ref.\cite{fleck} the effective mass of the nucleon at each
radius was computed by removing the average value of the square of the
momentum of the three quarks, computed in the bag at each radius. This
gives a very strong field dependence which reduces the vector potential
needed to reproduce the correct saturation properties of nuclear matter.
In Ref.\cite{finite} we studied the relativistic oscillator in an
external field and found that the field dependence of the c.m.
correction was, in fact, quite small. Therefore we have not
followed the precription of Ref.\cite{fleck}, but instead used a 
phenomenological c.m.
correction to the bag energy of the form $-z_0/R_B$, which is not
strongly dependent on the applied fields. As a consequence the vector
potential in this work tends to be a little bigger (and the nucleon
effective mass a little smaller) than in earlier work
\cite{hadrons,wilk,fleck,st1}.

\subsection{Quantization of the Motion of the Nucleon}

Having obtained the expressions for the energy and 
momentum of the bound, classical nucleon we can then quantize its
motion. In many ways the simplest quantization procedure would be to
set $\vec{P}\to -i\vec{\nabla}_R$ in a non-relativistic expansion of the
energy. There is then a small ambiguity over the ordering of
$\vec{\nabla}_R$ and $M_N^{\star}(\vec{R})$ which is discussed in detail in
Ref.\cite{finite}. An alternative procedure, which is designed to
clarify the connection to QHD, is to quantize using the Dirac equation
for the nucleon. In this case, the idea is to write a relativistic
Lagrangian which gives equivalent expressions for the nucleon energy and
momentum in mean-field approximation. This Lagrangian density is
\bge
{\cal L}=i\overline{\psi}\gamma \cdot \partial\psi-
M_N\overline{\psi}\psi
+g_\sigma(\hat{\sigma})\hat{\sigma}\overline{\psi}\psi
-g_\omega\hat{\omega}^\mu\overline{\psi}\gamma_\mu\psi+{\cal
L}_{mesons},
\label{6}
\ene
and clearly the only difference from QHD lies in the fact that the
internal
structure of the nucleon has forced a  (known) dependence of the scalar
meson-nucleon coupling constant, $g_\sigma(\hat{\sigma})$
on the scalar field itself. In terms of this coupling constant the
nucleon effective mass is
\bge
M_N^{\star}(\hat{\sigma})=M_N-g_\sigma(\hat{\sigma})\hat{\sigma}.
\label{7}
\ene

\section{Self-consistent Field Equations}

In the mean field approximation, the meson field operators in
Eq.(\ref{6}) are replaced by their time independent expectation values
in the ground state of the nucleus. As the resulting equations are in a
form which closely resembles the QHD equations in Hartree approximation
and one can relatively easily adapt existing computer programs to solve
the quark-meson coupling model, it seems worthwhile to summarise the field
equations here. For simplicity we shall retain only the $\sigma$ and
$\omega$ fields, although it is straightforward to generalise these
equations to include iso-vector mesons.

As explained above the nucleon satisfies the Dirac equation
\bge
( i\gamma \cdot \partial -M_N^{\star}(\sigma)-
g_\omega \gamma_0 \omega ) \psi = 0 ,
\label{8}
\ene
where the nucleon effective mass $M_N^{\star}(\sigma)$ is given by
Eq.(\ref{7}) and the scalar and vector fields satisfy
\bg
(-\nabla^2_r+m^2_\sigma)\sigma(\vec{r})&=& - \left(
\frac{\partial}{\partial\sigma}M_N^{\star}(\sigma) \right) \langle
A|\overline{\psi}\psi(\vec{r})|A \rangle, \label{9}\\
(-\nabla^2_r+m^2_\omega)\omega(\vec{r})&=&
g_\omega \langle A|\psi^\dagger\psi(\vec{r})|A \rangle .
\label{10}
\en

Note that the internal structure of the nucleon enters {\it only}
through the scalar field dependence of the scalar coupling constant. In
terms of the scalar charge of the nucleon
\bge
S(\vec{r}) = \int_{Bag} d\vec{u} \bar{q}(\vec{u}-\vec{r})
q(\vec{u}-\vec{r}),
\label{11}
\ene
(where $q$ is the quark wave function in the bound nucleon), which can
be expressed in closed form as
\bge
S(\vec{r}) =  \frac{\Omega_0/2 + m_q^{\star}R_B(\Omega_0
-1)}{\Omega_0(\Omega_0-1) +
m_q^{\star}R_B/2}, \label{12}
\ene
we can define $C(\sigma)$:  
\bge
C(\vec{r})=S(\vec{r})/S(\sigma=0).
\label{13}
\ene
Then for consistency $g_\sigma(\sigma)$ and $C(\sigma)$ must be related
by
\bg
C(\sigma)g_\sigma(\sigma=0)&=&
-\frac{\partial}{\partial\sigma}M_N^{\star}(\sigma) ,\nn \\
&=&\frac{\partial}{\partial\sigma}(g_\sigma(\sigma)\sigma). 
\label{14}
\en
It turns out that $C(\sigma)$ is well approximated by a linear form
\bge
C({\overline \sigma}) = 1 - a \times (g_{\sigma}{\overline \sigma}) ,
\label{15}
\ene
(where $g_{\sigma} \equiv g_{\sigma}(\sigma=0)$)
so that $C$ decreases by between 10 and 20\% between free space and the
density of normal nuclear matter\cite{finite}. Indeed, in this case one
can easily solve Eq.(\ref{14}) for $g_\sigma(\sigma)$, obtaining:
\bge
M^{\star}_N = M_N - \left[ 1 - \frac{a}{2} (g_\sigma {\overline \sigma})
\right] (g_\sigma {\overline \sigma}). 
\label{15'}
\ene
In conclusion, we note for completeness the
relation between the quark level coupling constants and those at the
nucleon level
\bge
g_\sigma =3 g_\sigma^q S(\sigma=0),\ \ g_\omega = 3 g_\omega^q.
\label{16}
\ene

\section{More on the Spin Orbit Force}

We saw earlier that the internal structure of the nucleon leads to a
spin orbit coupling to the (isoscalar) vector potential proportional to 
$1 - 2\mu_s$ (ignoring the small medium correction $\eta_s$). For the
$\rho$ meson we find the same expression but with the isovector nucleon
magnetic moment. Now in the isoscalar case it happens that $\mu_s$ is
approximately one so that $1 - 2\mu_s \approx -1$ which is what one
obtains directly from the non-relativistic reduction of the 
Dirac equation (\ref{8}). Thus one can simply use the Dirac equation
without any serious loss of accuracy.

On the other hand, in the isovector case one
has an isovector nucleon magnetic moment equal to 4.7 nuclear magnetons,
which is very far from unity and it appears that the Dirac formalism is
inappropriate. However, it is well known in the one-boson-exchange
models of the NN force, that the $\rho$ coupling to the nucleon has a
large anomalous piece, $f_{\rho} \bar{\psi}\sigma^{\mu\nu}\psi
\partial_{\nu}\rho_{\mu}$. In the mean field approximation 
such couplings can be ignored for nuclear matter because the meson field
is independent of position and time. The situation is rather different
in a finite nucleus, where
the time component of the vector field varies with radius. In fact, in
this case it is relatively straightforward to show that the
non-relativistic reduction of the Dirac equation, including an anomalous
coupling, gives a spin-orbit term equal to that derived in Eq.(\ref{8})
provided $f_{\rho}/g_{\rho}$ is chosen to be the isovector, anomalous
magnetic moment 3.7 $\equiv (\mu_p-1) - \mu_n$.

Clearly we could improve the accuracy of the treatment of the
$\omega$ too by adding a small anomalous, isoscalar term with
$f_{\omega}/g_{\omega} = -0.12$. It will be very interesting to extend
these considerations to other cases -- for example, the $\Lambda$ and 
$\Sigma$ hypernuclei. For an initial investigation of the masses of
hyperons in dense nuclear matter we refer to Ref.\cite{hadrons}. (Note,
however, that that work used the treatment of c.m. corrections to the
bag energy which we now believe to be inappropriate -- c.f. sect. 2,
above.)

\section{Initial Results}

\begin{figure}[htb]
\centering{\
\epsfig{angle=270,figure=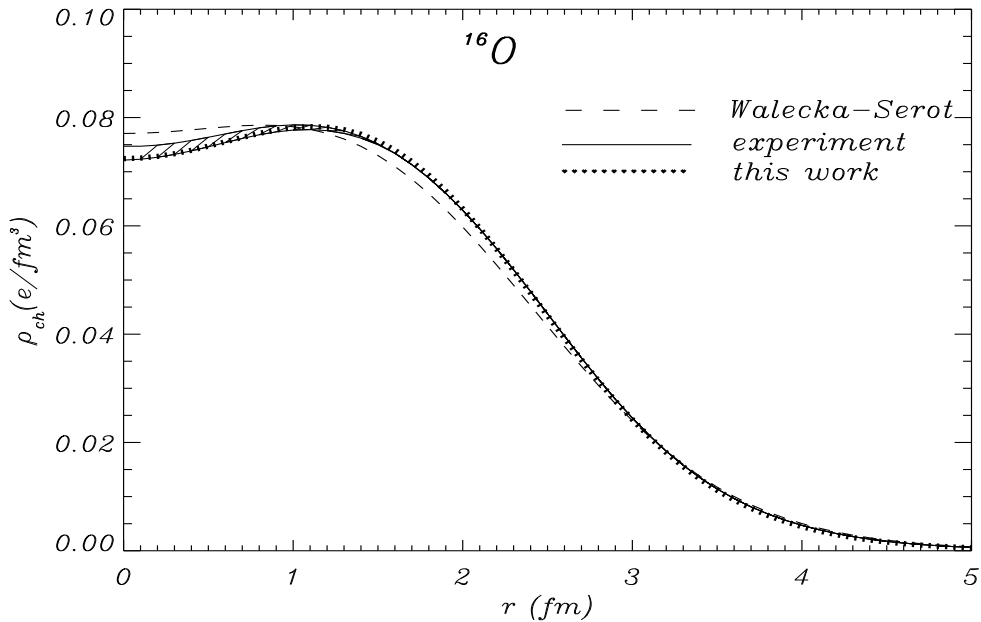,height=9cm,width=14cm,angle=0}
}
\parbox{130mm}{\caption{The charge density of $^{16}O$ in the
present model and QHD, compared with the experimental distribution.}
\label{fig:oxygen}}
\end{figure}

As an initial investigation of the application of the quark-meson coupling 
model
to finite nuclei we have considered the case of $^{16}O$. For the
protons one must, of course, include the central Coulomb repulsion.
The numerical calculation was
carried out using the techniques described by Walecka and
Serot~\cite{serot}.
The resulting charge density for $^{16}O$ is shown in
Fig.\ref{fig:oxygen}
(dotted curve) in comparison with the experimental
data~\cite{oxydata} (hatched area) and QHD~\cite{serot}.

The parameters used correspond to a free bag radius of 0.8 fm --
although this shrinks by about 2\% at nuclear matter density. As the
central density tended to be a little high in comparison with experiment
we increased the model dependent slope parameter, $a$ in Eq.(\ref{15}),
by about 10\% (above that calculated in the bag model) 
to obtain the results shown. The corresponding effect on
the saturation energy and density of nuclear matter was very small.
It is interesting to note, although the physical significance of the
observation is not at all clear, that if the quark mass was taken to be
around 300 MeV, rather than near zero (i.e. a constituent mass rather
than a current quark mass) the density of $^{16}O$ was just as good
without the need to adjust $a$ at all.

\section{Concluding Remarks}

Having made so much progress in the development of the quark-meson
coupling model there is a great deal of interest in exploring its
consequences. The obvious extensions of the work described here and in
Ref.\cite{finite} to heavier nuclei are already underway. In view of the
promising results for nuclear charge symmetry breaking and $\beta$-decay
obtained using local density approximation we are also keen to explore
these applications in a genuine finite nucleus. For hypernuclei the
natural extension (c.f. Ref.\cite{hadrons}) is to assume that the
$\sigma$ and $\omega$ mesons couple only to the non-strange
constituents. From our discussion of the spin orbit force in sect. 4 and
the fact that the spin of the $\Lambda$ is carried entirely by the
strange quark, one can easily see that the $\Lambda$ spin orbit force
will arise entirely from the Thomas precession term. As the scalar and
vector potentials tend to cancel in that term (c.f. Eq.(\ref{3})), this
means that the $\Lambda$ spin orbit force is very naturally suppressed
in this model -- as observed experimentally. It will be important to
follow this observation with quantitative results.

In view of the suggestion that vector meson masses may be substantially
lower in dense matter\cite{brown,expt} it will also be interesting to
repeat our earlier work\cite{hadrons} with the new treatment of the c.m.
correction -- i.e. with our larger scalar and vector fields. As a first
estimate, however, we can take the lesson of Ref.\cite{hadrons} that the
reduction in the mass scales with the number of non-strange quarks and
the result in the present model that the nucleon effective mass is of
the order of 600 MeV at $2.5\rho_0$ to estimate that at such densities
the effective mass of the $\rho$ meson should also be around 600 MeV.
This seems to be roughly the range needed to understand the current
experiments.

In terms of further theoretical development it will be interesting to
compare the present model with more phenomenological, non-linear
extensions of QHD -- as reviewed recently in Ref.\cite{delf}. We would
also like to consider the replacement of the effective $\sigma$ meson
exchange by two-pion-exchange within a chiral quark model such as the
cloudy bag\cite{cbm}. Finally one would also like to find ways to
replace at least some of the repulsion associated with $\omega$ exchange
by nucelon overlap with quark and gluon exchange. 

\section*{Acknowlegements}

This work was supported in part by the Australian Research Council.


\begin{thebibliography}{99}
%
\bibitem{guichon} P.~A.~M.~Guichon, Phys. Lett. {\bf B200} (1988) 235.
%
\bibitem{walec} J.~D.~Walecka, Ann. Phys. (N.Y.) {\bf 83} (1974) 497;
\\
S.~A.~Chin, Ann. Phys. (N.Y.) {\bf 108} (1977) 301.
%
\bibitem{serot} B.~D.~Serot and J.~D.~Walecka, Adv. Nucl. Phys.
{\bf 16} (1986) 1.
%
\bibitem{hadrons} K.~Saito and A.~W.~Thomas, Phys. Rev. {\bf C51} (1995)
2757.
%
%
\bibitem{ons} K.~Saito and A.~W.~Thomas, Phys. Lett. {\bf B335} (1994)
17.
%
\bibitem{wilk} K.~Saito and A.~W.~Thomas, Phys. Lett. {\bf B363} (1995)
157.
%
\bi{finite} P.~A.~M.~Guichon et al., nucl-th/9509034, to appear in Nucl.
Phys. {\bf A} (1996).
%
\bibitem{jackson} J.D. Jackson, Classical Electrodynamics, J. Wilew \&
Sons, New York (1967) 364.
%
\bibitem{goldstein} H. Goldstein, Classical Mechanics, Addison-Wesley,
Menlo Park Ca USA (1968) 212.
%
\bibitem{fleck} S.~Fleck {\it et al.} Nucl. Phys. {\bf A510} (1990)
731.
%
\bi{st1} K.~Saito and A.~W.~Thomas, Phys. Lett. {\bf B327} (1994) 9.
%
\bibitem{oxydata} I.~Sick and J.~S.~McCarthy, Nucl. Phys. {\bf A150}
(1970) 631.
%
\bi{brown} G.~Q.~Li, C.~M.~Ko and G.~E.~Brown, Phys. Rev. Lett. {\bf 75}
(1995) 4007.
%
\bi{expt} G.~Agakichiev et al. (CERES Collaboration), Phys. Rev. Lett.
{\bf 75} (1995) 1272; M.~Masera et al. (HELIOS-3 Collaboration), Nucl.
Phys. {\bf A590} (1995) 93c.
%
\bi{delf} A. Delfino et al., nucl-th/9602004 (1996).
%
\bi{cbm} A.W. Thomas, Adv. Nucl. Phys. {\bf 13} (1984) 1.
%
\end{thebibliography}
\end{document}